\documentclass[]{aa}
\usepackage{graphicx,psfrag,amsmath}
\usepackage{natbib}
\bibpunct{(}{)}{;}{a}{}{,}
\begin{document}

   	\title{Vertical     distribution      of     Galactic     disk
	stars:\thanks{Based on observations  taken at the Observatoire
	de  Haute  Provence (OHP)  (France),  operated  by the  French
	CNRS.}} 
	\subtitle{II.  The  surface mass  density  in  the  Galactic  plane  }
	\titlerunning{ The surface mass density in the Galactic plane}

   \author{A. Siebert\inst{1}, O. Bienaym\'e\inst{1}, C. Soubiran\inst{2}}
	\authorrunning{Siebert et al.}
   \offprints{siebert@astro.u-strasbg.fr}
%
   \institute{Observatoire  Astronomique   de  Strasbourg,  UMR  7550,
              Universit\'e  Louis  Pasteur,  Strasbourg,  France  \and
              Observatoire  Aquitain des  Sciences  de l'Univers,  UMR
              5804, 2 rue de l'Observatoire, 33270 Floirac, France}
   \date{Received \today  / Accepted }
   \abstract{ High  resolution spectra data  of red clump stars  towards the
NGP have been  obtained with the high resolution  spectrograph Elodie at OHP
for  Tycho-2 selected stars.   Combined with  Hipparcos local  analogues, we
determine  both the gravitational  force law  perpendicular to  the Galactic
plane,  and the total  surface mass  density and  thickness of  the Galactic
disk.  The surface mass density  of the Galactic disk within 800\,pc derived
from  this   analysis  is  $\Sigma$   ($|z|<$800  \,  $\rm{pc})$  =   76  \,
$\mathrm{M}_{\odot}   \mathrm{pc}^{-2}$   and,   removing  the   dark   halo
contribution,   the   total  disk   mass   density  is   $\Sigma_0$\,=\,67\,
$\mathrm{M}_{\odot} \mathrm{pc}^{-2}$ at solar radius.  The thickness of the
total disk mass distribution is  dynamically measured for the first time and
is found  to be 390$^{+330}_{-120}$\,pc  in relative agreement with  the old
stellar  disk scaleheight.   All  {\it dynamical}  evidences concerning  the
structure of  the disk (its local  volume density --i.e.   the Oort limit--,
its  surface density  and its  thickness)  are compatible  with our  current
knowledge of the corresponding stellar disk properties.

}
   \maketitle
%
%
\section{Introduction}
We present a new dynamical determination of the Galactic potential and $K_z$
force  perpendicular to  the Galactic  plane.  We  measure the  surface mass
density of the Galactic disk (i.e. the total amount of disk mass in a column
perpendicular to the Galactic plane)  and also the thickness of the vertical
mass distribution of  the disk at the Sun.   They are fundamental parameters
of the Galaxy,  determining the disk contribution to  the Galactic potential
and to the rotation curve.  They  also give valuable constraints to the disk
properties.  The  ``$K_z$ problem'' is traditional in  Galactic dynamics and
subsequent derivation  of the total  mass density $\rho_0$ (the  Oort limit)
has a long history  begining with \citet{K22} and \cite{O32}.  Comprehensive
review may  be found in \citet{KL86}  and in \cite{K95}.   Most recent works
may be  found in  \citet{CC98a} and in  \citet{HF00} and  references within.
They favor a low Oort limit $\rho$(z=0)\,=\,0.076--0.10\,$\mathrm{M}_{\odot}
\mathrm{pc}^{-3}$.  Previous surface  mass density determinations (including
only  disk  contribution  and  removing  the local  halo  contribution)  are
$\Sigma_0$=\,52$\pm13\,\mathrm{M}_{\odot}\mathrm{pc}^{-2}$      \citep{FF94},
$\Sigma_0$\,=\,48$\pm9\,\mathrm{M}_{\odot} \mathrm{pc}^{-2}$, \citep{KG91}.

Here, to  investigate the vertical  structure and potential of  the Galactic
disk,  we have  observed a  sample  of red  clump giants  towards the  North
Galactic  Pole  with  a   high  resolution  spectrograph  (Elodie  at  OHP).
Observations and properties  of the sample are given in the  paper I of this
series  \citep{SBS02}  where  we  detail  the  measurements  of  abundances,
absolute  magnitudes   and  radial  velocities,  and   describe  the  sample
properties.  Atmospheric  parameters and  absolute magnitudes of  stars have
been obtained with  the TGMET method \citep{KS98} by  comparing the observed
spectra  to the TGMET  library of  similar spectra  of Hipparcos  stars with
known characteristics.  The method is applied in Paper I \citep{SBS02}.  The
library is  built from  the ELODIE database  described in  \citet{PS01}. The
main  advantage  of selecting  red  clump  stars  is that  their  luminosity
function  is  sharply  defined  in  a small  magnitude  interval  minimizing
completeness  bias (Malmquist  or Lutz-Kelker),  making more  accurate their
corrections.   This remote  sample has  been  analysed combined  to a  local
sample of similar Hipparcos stars in a sphere of 125\,pc around the Sun.

In Sect.\,2, we describe the  red clump sample definition extracted from the
general sample  of bright red  stars observed at  OHP (see Paper I)  and the
correction for the Lutz-Kelker  bias for the associated Hipparcos subsample.
The method to determine the  potential $\Phi(z)$ and vertical force $K_z$ is
classical  but calls for  some algebra  because the  NGP samples  cover wide
solid angles on  the sky (Sect.\,3).  Results are  presented in Sect.\,4: we
confirm with  this independent  stellar sample previous  findings concerning
the  disk  surface mass  density  and  we  present the  first  observational
constrain on  the total thickness of  the mass distribution  of the Galactic
disk.

    \begin{figure*}[hbtp]
     \centering
       \includegraphics[width=6.5cm]{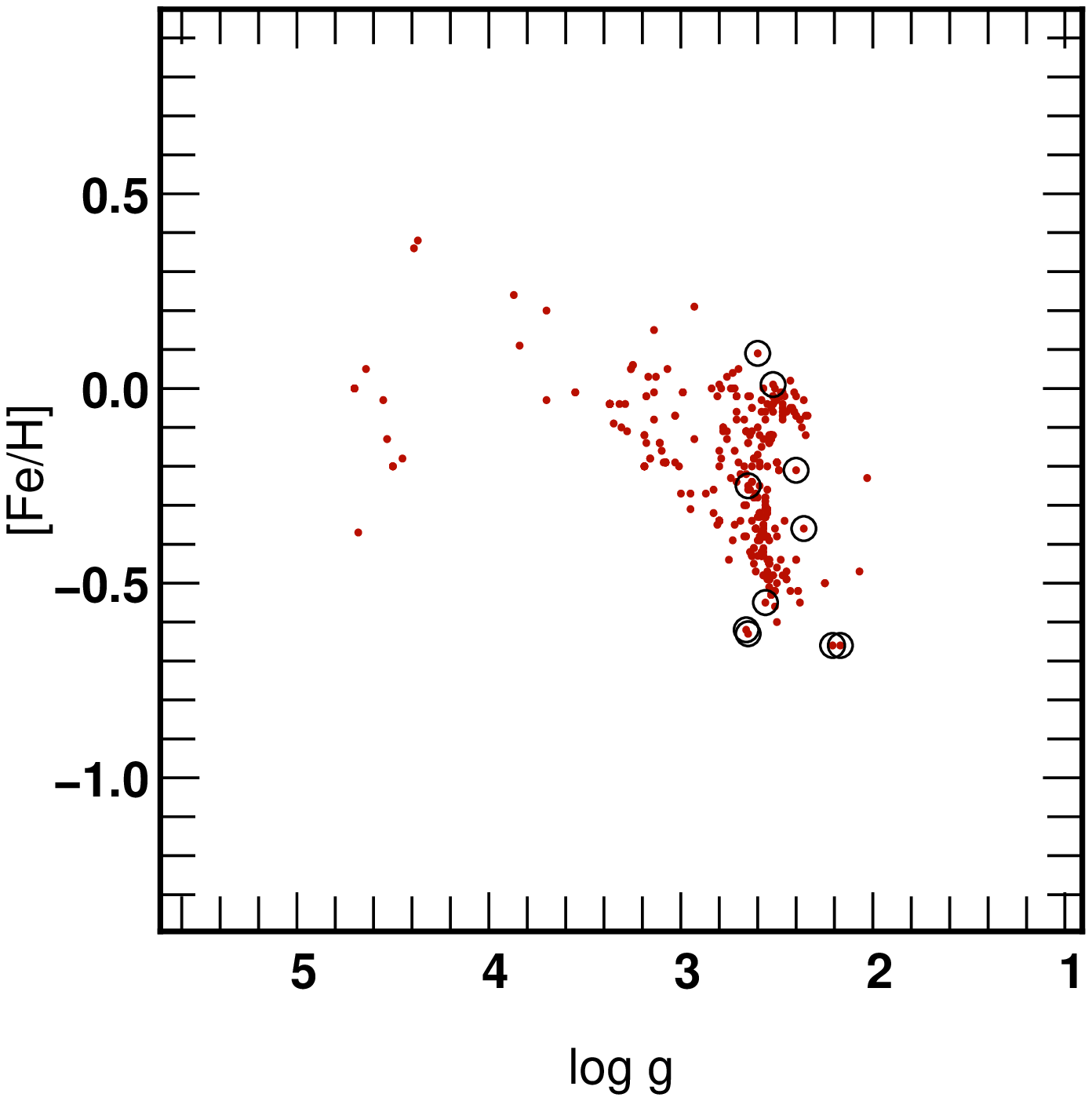}
       \includegraphics[width=6.5cm]{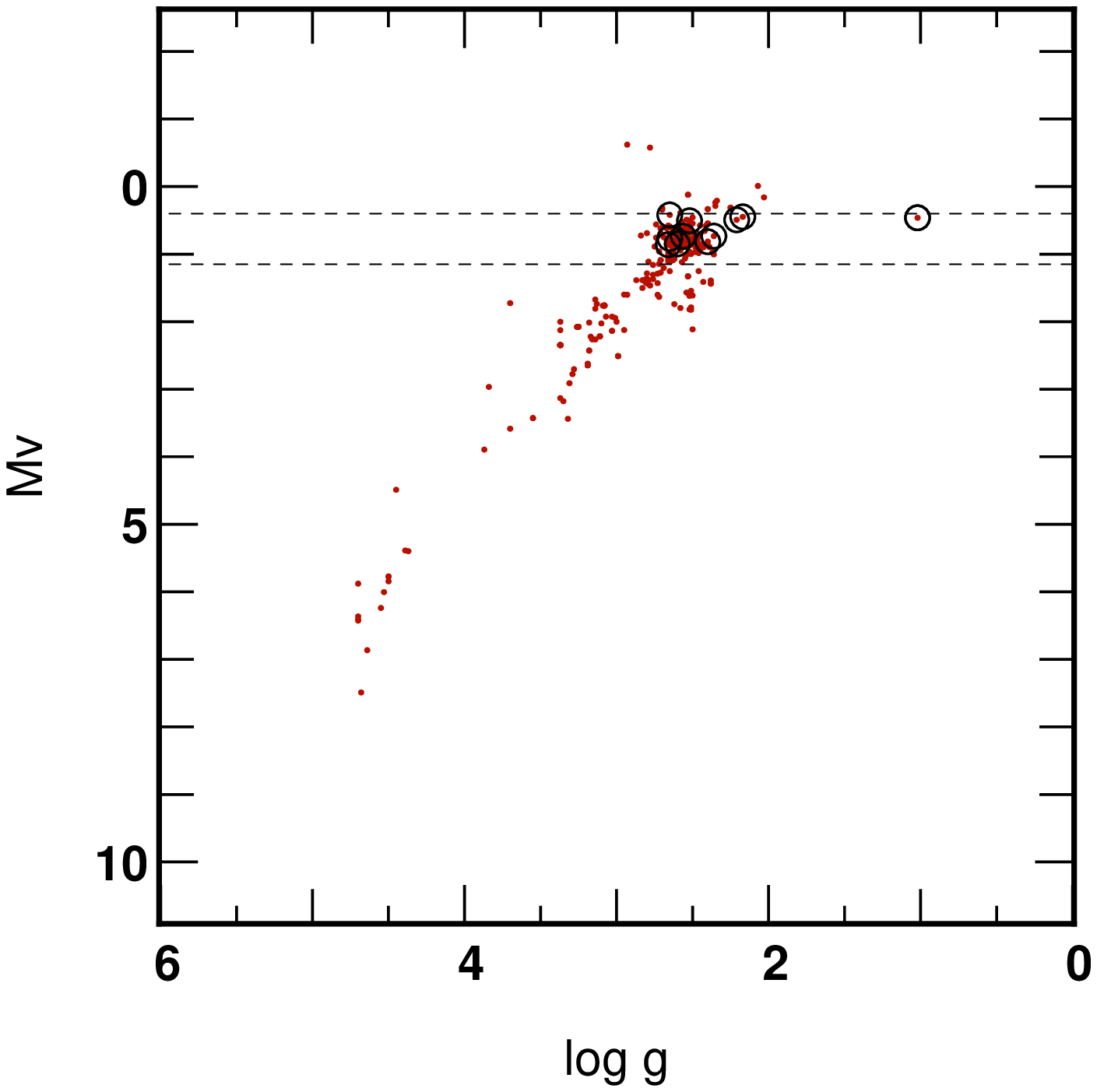}
       \includegraphics[width=6.5cm]{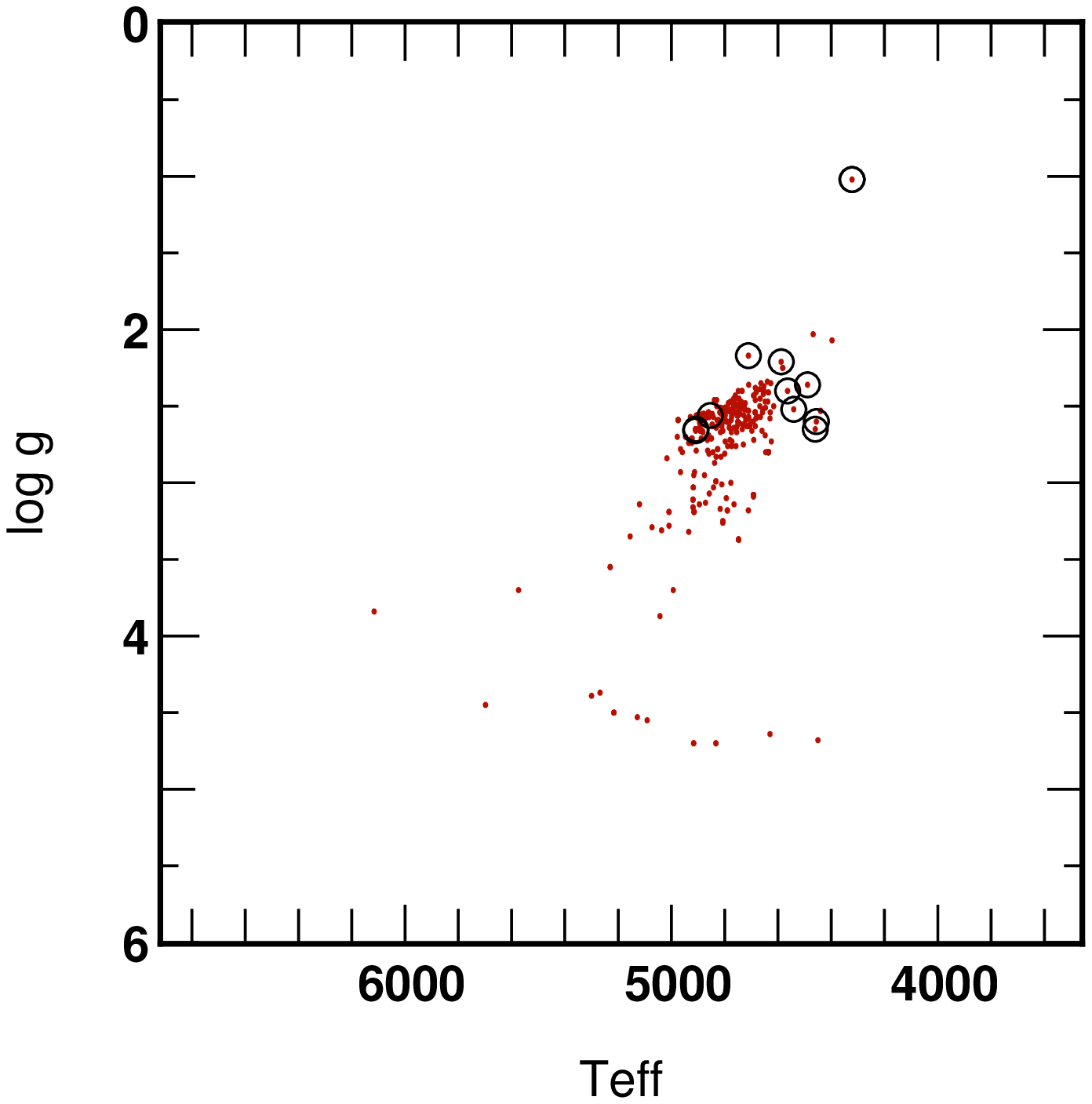}
       \includegraphics[width=6.5cm]{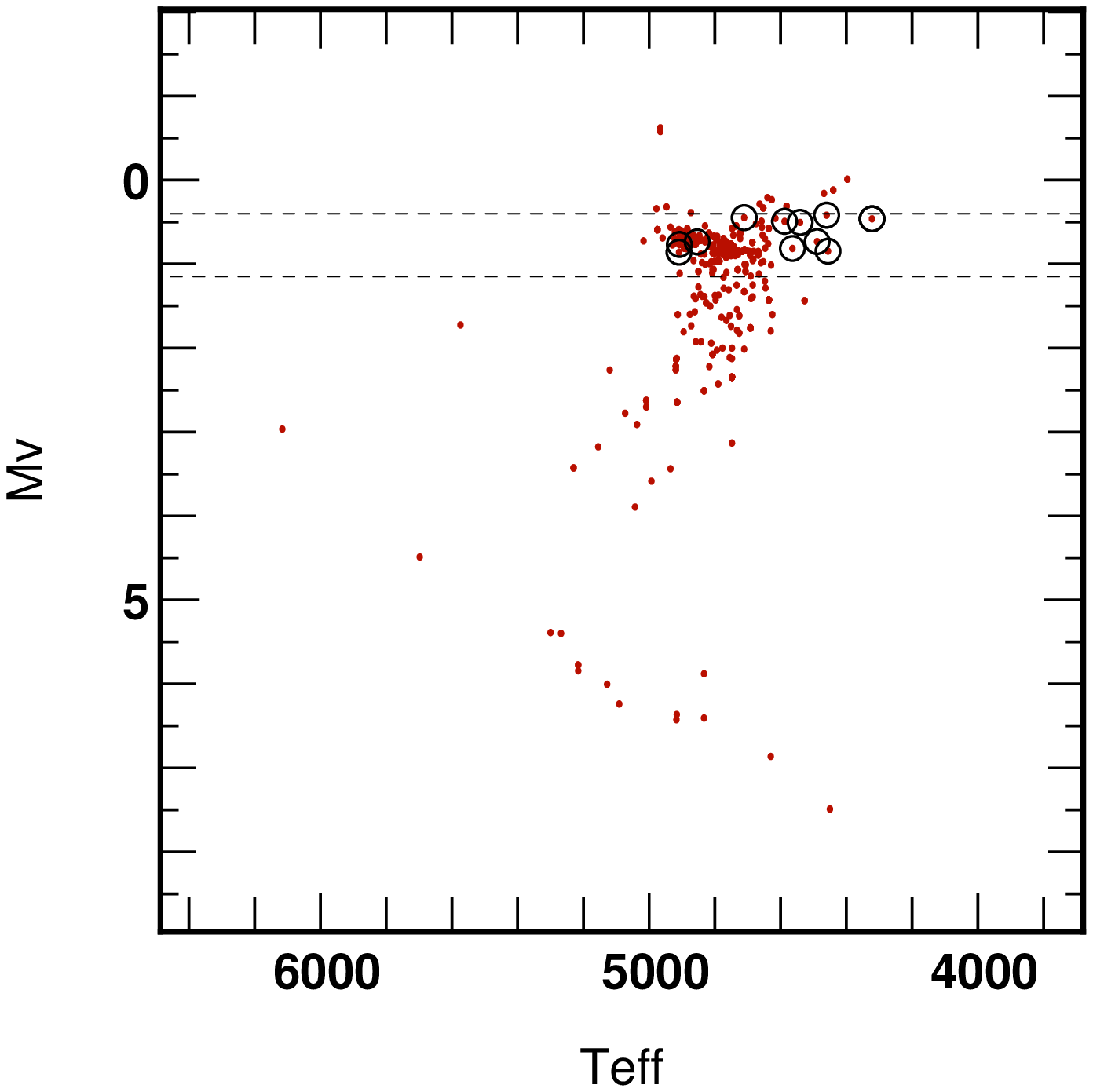}
       \caption{Properties of the observed stars.  Horizontal dashed lines
       indicate our selection procedure for the red giant clump sample. 
       Circles  denote stars belonging to our $M_{\rm V}$ selection
       that  are probably not clump stars (see text).}
    \label{fig:selection1}
    \end{figure*}
%
\section{The giant clump samples}
\label{s:giantclump}
Our  sample  consists  of  stars  selected  out  of  the  Tycho-2  catalogue
\citep{HIP} and is described in Paper I \citep{SBS02}.  A selection in $B-V$
colour  has been  applied to  increase the  number of  red clump  stars with
respect to dwarfs, subgiants, RGB and  AGB stars.  We have chosen the colour
interval 0.9\,$<B-V<$\,1.1 that optimizes  the detection of red clump stars.
The  lower $B-V$  limit at  0.9 rejects  the brightest  main  sequence stars
($M_{\textrm  v}<5$),  some subgiants  and  also  most  giants of  the  blue
horizontal branch  (the "clump stars"  with the lowest  metallicities).  The
upper $B-V$ limit at 1.1 rejects most of RGB and AGB stars.\\
The selection  criteria maximise the number  of clump stars  with respect to
dwarfs and  red giants along  the ascending branch  but remain a  mixture of
those  three  types  of   objects.   Our  stellar  parameters  determination
(absolute magnitude) allows to separate red clump stars from dwarfs and from
other  giant stars.  We combine  our remote  sample with  a  local Hipparcos
sample  built  with  the  same   criteria  based  on  colours  and  absolute
magnitudes.         This        procedure        is       sketched        in
subsection~\ref{s:NGPsamples}. The  distance limited sample  is built, using
the Hipparcos catalog.  This sample is affected by the Lutz-Kelker bias, its
composition   as    well   as   the    bias   correction   are    given   in
subsection~\ref{s:HIPsample}.

\subsection{The NGP cone samples} 
\label{s:NGPsamples}
In this study we are interested  in the bright stellar population defined by
the  red clump  stars allowing  to probe  stellar populations  far  from the
mid-plane.

Red  clump  stars  were selected  on  an  absolute  magnitude and  a  colour
criteria.  Those  criteria are defined to  maximize the number  of stars but
within a small absolute magnitude  interval.  The reason is to minimize {\it
a priori} the Malmquist and/or Lutz-Kelker bias.  Moreover, isolating clump
stars allows to define a well defined sample in term of stellar evolution.
On  Fig.~\ref{fig:selection1},  red clump  stars  appear  as an  overdensity
mainly in the  [$M_{\rm V}$, $\log g$] and [$\log  g$, $T_{\rm eff}$] planes
enabling us to perform a reliable selection.  An absolute magnitude interval
is set, using the selection of the overdensity, to retrieve a maximum of red
clump stars and  a minimum of background stars. The  horizontal lines on the
two right panels correspond  to $M_{\rm V}$\,=\,0.4 and $M_{\rm V}$\,=\,1.15
which is the best compromise between the total number of red clump stars and
the number  of background objects in  the resulting sample.   Circles on the
four panels show stars that do  not belong to the overdensity (i.e.  the red
clump),  but belong  to the  color  and absolute  magnitude selection.   The
number of stars in the sample  rises from 221 (overdensity selection) to 232
(absolute  magnitude criteria)  showing that  a small  fraction  ($<5\%$) of
background stars are included.

\begin{figure}[hbtp]
    \begin{center}
    \includegraphics[width=6cm]{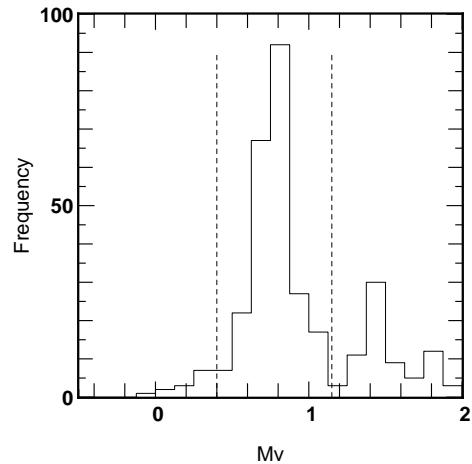}
    \caption{Absolute  magnitude  histogram  of  the observed NGP samples.
    The vertical lines define the  selection  of the  red clump  stars
    around   the   central   peak.}
    \label{fig:selection2}
    \end{center}
    \end{figure}

The  histogram on Fig.~\ref{fig:selection2}  draws the  presence of  the red
clump  as a  narrow peak  at  $M_{\rm v}$=0.8.   The two  vertical lines  at
$M_{\rm v}=0.4$  and $1.15$  are our selection  interval.  Only a  few other
type of giants contribute to the background.

This sample  (two fields towards the  NGP) is separated  in four subsamples,
two distance-complete  samples and two  magnitude complete samples,  one for
each field.  The resulting  number of  stars in each  subsample is  given in
Tab.~\ref{t:samples}, to be compared to the 537 similar stars extracted from
the Hipparcos catalogue (next section).

\begin{table*}
\caption{\label{t:samples}
Description of the north Galactic pole samples }
\center
\begin{tabular}{l c c c c}\\
\hline
\hline
Field & d$_{low}$ & d$_{up}$ & number of stars & surface (square deg)\\
\hline
Field 1 & 226 & 616 & 85 & 309.4\\
Field 2 & 226 & 405 & 49 & 410.1\\
\hline
Field & V$_{low}$ & V$_{up}$ & number of stars & surface (square deg)\\
\hline
Field 1 & 7.2 & 10.1 & 128 & 309.4\\
Filed 2 & 7.2 & 9.2 & 73 &410.1\\
\hline
\end{tabular}
\end{table*}
    \subsection{The Hipparcos sphere sample} 
    \label{s:HIPsample}
    A  local  sample of  clump  giants has  been  drawn  from the  Hipparcos
    catalogue \citep{HIP} which provides  $B$ and $V$ magnitudes, parallaxes
    and proper motions.  This sample  is corrected from the Lutz-Kelker bias
    using     \citet{TLC77}     method      which     is     outlined     in
    section~\ref{s:LKcorrection}. The selection criteria used for extracting
    red  clump giants  is transposed  to the  Hipparcos sample  in  order to
    ensure  the  homogeneity  of  the  different samples.   The  sample  was
    completed  with   radial  velocities  from  the   literature,  the  full
    description     of      its     construction     is      outlined     in
    section~\ref{s:constructionHIP}.

     \subsubsection{Lutz-Kelker bias and corrections}
     \label{s:LKcorrection}
	Conversion  of parallaxes  into distances  introduces a  bias, stars
	appearing statistically  further away  than they really  are.  Since
	the proposed correction of  this bias by \citet{LK73}, other methods
	have  been  proposed  \citep{TLC77,Ar99}.   To  adjust  the  overall
	distribution and also  the individual values, we use  the process of
	\citet{TLC77} that allows us to correct the magnitude bias according
	to the precise shape of the LF.  The same method has been applied by
	\citet{Gi98} to analyse Hipparcos giants.  Here we derive the formal
	correction  for  distances.   The  later corrections  enable  us  to
	defined a volume limited sample, free from Lutz-Kelker bias.

	Relating the observed absolute magnitude $M_{\mathrm{obs}}$ to
	the true absolute  magnitude $M_{\mathrm{true}}$, we can write
	the unbiased estimator of $M_\mathrm{true}$ as:
        $$<\!M_{\mathrm{obs}} + \Delta M\!>\,=\,M_{\mathrm{true}} \mbox{,}$$
	where $\Delta M$ is the corrective term. It follows that
	\begin{eqnarray}
        <\!\Delta M\!> & = & <\!M_{\mathrm{true}}-M_{\mathrm{obs}}\!>
        \nonumber \\
                       & = & \frac{\int
        (M_{\mathrm{true}}  - M_{\mathrm{obs}}  ) {\rm d} {\cal P}(M_{\mathrm{true}}  |
        M_{\mathrm{obs}})  }  {\int {\rm d}{\cal P}(M_{\mathrm{true}}|M_{\mathrm{obs}})}
        \mbox{.}
        \label{e:LKcorindiv} 
        \end{eqnarray}
	where  ${\cal P}(M_{\mathrm{true}}|M_{\mathrm{obs}})$  is the  conditional
	probability  that the absolute  magnitude is  $M_{\mathrm{true}}$ if
	the    observed   one   is    $M_\mathrm{obs}$.    We    derive
	${\rm d}
	{\cal P}(M_{\mathrm{true}}|M_{\mathrm{obs}})$  which  is  proportional  to
	${\cal P}(\pi_{\mathrm{true}}|\pi_{\mathrm{obs}}){\rm d}\pi_{\mathrm{true}}$
	\citep{TLC77}.   ${\cal P}(\pi_{\mathrm{true}}|\pi_{\mathrm{obs}})$ is the
	conditional probability that $\pi_{\mathrm{obs}}$ being the observed
	value, the real parallax is $\pi_{\mathrm{true}}$.
	Under the assumption of gaussian  errors on the parallaxes and for a
	uniform density of stars, this term is given by
        $${\cal    P}(\pi_{\mathrm{true}}|\pi_{\mathrm{obs}})    =    \frac{
\pi_{\mathrm{true}}^{-4}
\Phi(M_{\mathrm{true}})}{\sqrt{2\pi}\sigma_{\pi_{\mathrm{true}}}}\exp\left[
-
\frac{(\pi_{\mathrm{obs}}-\pi_{\mathrm{true}})^2}{2\sigma_{\pi_{\mathrm{true}}}^2}
\right] \mbox{,}$$
	where $\Phi(M_{\mathrm{true}})$ is  the luminosity function of
	the   studied  tracer  population.    We  assume   a  gaussian
	distribution  for  red clump  stars  with  mean $<\!M\!>$  and
	standard  deviation  $\sigma_{RC}$.  Therefore  $\Phi(M)$
	writes
	\begin{eqnarray} 
        \Phi(M) & = & a_{0} + a_{1}M + a_{2}M^{2} \nonumber\\
                & + & \frac{\mathrm{A}}{\sqrt{2
	\pi}   \sigma_{RC} }\exp  \left[-  \frac{(M  -  <\!M\!>)^2}{2
	\sigma_{RC}^2}\right] \mbox{,}
	\label{e:LF}
	\end{eqnarray}
	where the second  order polynomial represents  the contribution of
	other giants.

	Calling   
        $$G_{\pi_{\mathrm{obs}}}(\pi_{\mathrm{true}})   =  \frac{\pi_{\mathrm{true}}^{-4}
	\Phi(M_{\mathrm{true}})}{\sqrt{2\pi}\sigma_{\pi_{\mathrm{true}}}}\exp\left[ -
	\frac{(\pi_{\mathrm{obs}}-\pi_{\mathrm{true}})^2}{2\sigma_{\pi_{\mathrm{true}}^2 }}
        \right]$$ 
	$$M_{\mathrm{true}} \approx 5 \log(\pi_{\mathrm{true}}) +m -5$$
	$m$ being  the apparent  magnitude of the  star (noting that  we are
	dominated by  errors on the  parallaxes and that errors  on apparent
	magnitude are negligible), and
        $$\left(  M_{\mathrm{true}}  -   M_{\mathrm{obs}}  \right)  =
	5 \log  \left( \frac{\pi_{\mathrm{true}}}{\pi_{\mathrm{obs}}} \right)$$
	the correction given by Eq.~\ref{e:LKcorindiv} rewrites
        \begin{equation}                                            
        <\!\Delta M\!>\,=\,\frac{\int_{0}^{\infty}G_{\pi_{\mathrm{obs}}}(\pi_{\mathrm{true}})
        5 \log \left( \frac{\pi_{\mathrm{true}}}{\pi_{\mathrm{obs}}} \right)
        {\rm d}\pi_{\mathrm{true}}}{\int_{0}^{\infty}G_{\pi_{\mathrm{obs}}}(\pi_{\mathrm{true}})
        {\rm d}\pi_{\mathrm{true}}} \mbox{.}  
        \label{e:Mvcor} 
        \end{equation}
	Using the same approach, we derive the mean correction to be applied
        to the distances. This correction is given by
	\begin{equation}
	<\!\Delta d\!>\,=\,
	\frac{\int_{0}^{\infty}G_{\pi_{\mathrm{obs}}}(\pi_{\mathrm{true}})
	\left( \frac{1}{\pi_{\mathrm{true}}}-\frac{1}{\pi_{\mathrm{obs}}} \right)
        {\rm d}\pi_{\mathrm{true}}}{\int_{0}^{\infty}G_{\pi_{\mathrm{obs}}}(\pi_{\mathrm{true}})
        {\rm d}\pi_{\mathrm{true}}} \mbox{.} 
	\label{e:dcor}
	\end{equation}
	Equations~\ref{e:Mvcor} and~\ref{e:dcor}  allows us to  correct on a
	statistical basis the individual values of the distance and absolute
	magnitude, given the luminosity function of our sample.
      \subsubsection{Red Clump luminosity function} 
      \label{s:RCLF}
	To correct  the observed magnitudes  from Lutz-Kelker bias,  we must
	determine the  luminosity function  of the tracer  population.  This
	luminosity function is  obtained using a nearby sample  built out of
	the  Hipparcos catalogue.  We  select all  stars closer  than 50\,pc
	(i.e.  parallaxes $>$ 20 mas), with absolute magnitudes in the range
	[0.2,\,1.3]  and with  $B\,-\,V$ colours  in the  range [0.9,\,1.1].
	This  selection  gives a  distribution  that  can  be considered  as
	unbiased from  errors on  parallaxes.  The observed  distribution of
	absolute magnitudes is therefore  close to the true distribution and
	we  estimate it  by  a least-square  fit  of the  function given  by
	Eq.~\ref{e:LF}.

\begin{figure}[hbtp]
	\centering
	\includegraphics[width=6cm]{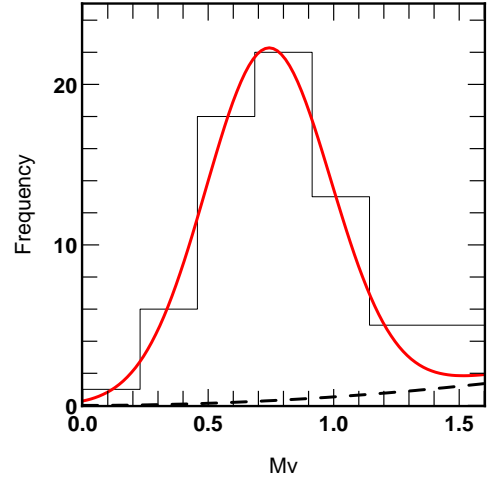}
	\caption{Distribution  of absolute  magnitudes for  the  clump stars
	closer than 50\,pc from the sun.  The observed distribution has been
	fitted  using Eq.~\ref{e:LF}  giving $<\!M_{\rm  v}\!>\,=\,0.74$ and
	$\sigma_{M_{\rm v}}\,=\,0.25$  for the local clump.  The dashed line
	shows the contribution of field giants to the sample.}
	\label{fig:RCLF}
	\end{figure}

	Figure~\ref{fig:RCLF} shows  the result of the  least-square fit and
	the   observed  distribution.   The   local  clump   parameters  are
	$<\!M_{\rm   v}\!>\,=\,0.74$,   $\sigma_{M_{\rm  v}}\,=\,0.25$   and
	$A=13.61$.   The   other  giants,  modelled  by   the  second  order
	polynomial (  best fit parameters $a_{0} \sim  10^{-4}$, $a_{1} \sim
	10^{-9}$  and  $a_{2}=0.0531$) give  a  contribution  to the  sample
	smaller than $2\%$.

      \subsubsection{Construction of the sample} 
      \label{s:constructionHIP}
The local  sphere sample is built  using Hipparcos data where  we select all
stars with parallaxes larger than 5 mas (i.e.  stars closer than 200\,pc) in
the same colour interval as the two NGP samples.

\begin{figure}[hbtp] 
\begin{center}
\includegraphics[width=6cm]{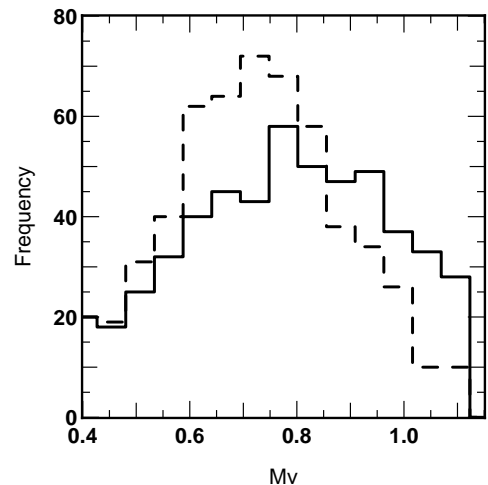}
\caption{Observed luminosity  distribution (continuous line)  versus the
    Lutz-Kelker bias corrected luminosity distribution (dashed line).}
\label{fig:LFcor} 
\end{center} 
\end{figure}

Then  we  correct the  sample  from Lutz-Kelker  bias  and  we compare  both
corrections,    first    using    the    absolute    magnitude    correction
(Eq.~\ref{e:Mvcor})    and   second    using    the   distance    correction
(Eq.~\ref{e:dcor}).  The mean absolute magnitude correction of the sample is
$-0.09$  mag with  a  standard deviation  equal  to 0.13  mag,  and for  the
distances the mean correction is  $-5.7$\,pc with a standard deviation equal
to  23.5\,pc.  Figure~\ref{fig:LFcor}  shows the  effect of  the Lutz-Kelker
bias on the luminosity function before and after correction.  The continuous
line  gives the  observed distribution  in absolute  magnitude,  whereas the
dashed line shows the corrected  luminosity function.  The sample defined by
stars  closer  than 125\,pc  (537  stars)  was  then established  using  the
corrected  distances.  The  consistency has  been checked  by  comparing the
distance  obtained after  correction with  Eq.~\ref{e:dcor} to  the distance
derived using the absolute magnitude  correction.  The difference in the two
distances being of  the order of 0.2\,pc for the sample,  this ensures us of
the reliability of the correction.

We  use  the  \citet{BB00}  catalogue  that  provides  us  with  the  radial
velocities.  We complete  the sample with radial velocities  from the Simbad
database when  measurements are existing.   This leads to a  nearly complete
(98\%) sample  of 526 stars  (out of 537  stars). Among those 526  stars, 27
have radial velocities  from Simbad whereas 499 have  radial velocities from
\citet{BB00}.

\begin{figure}[hbtp] 
\begin{center}
\includegraphics[width=6cm]{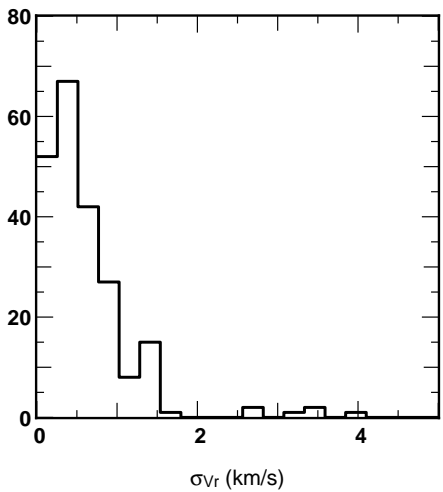}
\caption{Distribution of standard errors for the stars of the Hipparcos sample
with given $\sigma_{Vr}$ in \citet{BB00} (224 stars). 6 stars have errors
larger than 5 km/s and are not within the limits of the plot.}
\label{fig:sVrBB} 
\end{center} 
\end{figure}

Measurements errors for the radial velocities are given in two distinct ways
in \citet{BB00},  either a value  for $\sigma_{Vr}$ is given  when available
(224 stars)  or a quality index (A  to D) for stars  whose radial velocities
are    obtained   from    a    prism   objective    survey   (275    stars).
Figure~\ref{fig:sVrBB}  shows  the  distribution  of errors  of  the  radial
velocity for stars with  known $\sigma_{Vr}$ while Tab.~\ref{t:BB} gives the
distribution of the quality index for the stars with no $\sigma_{Vr}$.

\begin{table}
\caption{\label{t:BB}
Distribution of the quality index for radial velocity of the stars of our
sample in \citet{BB00}.}
\center
\begin{tabular}{c r c l c}\\
\hline
\hline
Quality Index &  & $\sigma_{Vr}$ & (km/s) & number of stars\\
\hline
A &            & $\sigma_{Vr}$ & $\leq$ 2.5 & 74\\
B & 2.5  $<$ & $\sigma_{Vr}$ & $\leq$ 5.0 & 170\\
C & 5.0  $<$ & $\sigma_{Vr}$ & $\leq$ 10.0& 23\\
D & 10.0 $\le$ & $\sigma_{Vr}$ &            & 8 \\
\hline
\end{tabular}
\end{table}

%
%
\section{The force perpendicular to the Galactic plane and the  disk mass distribution}
\label{s:model}
Our  first assumption  is that  stellar  population distributions  are in  a
stationary  state  with  well  mixed  distribution  functions  according  to
coordinate   z  (distance  from   the  Galactic   plane)  and   w  (vertical
velocity). The  second assumption  is that the  stellar motion  separates in
vertical and  horizontal motions (separable potential),  so the distribution
function $f(z,w)$  of a tracer population  depends only on  the potential of
the disk mass and  on the kinetic energy of stars.  For  instance, it may be
written as:
\begin{equation}
f(z,w)=\sum_{i=1}^{N}\frac{\rho_{i}(0)}{\sqrt{2 \pi}
\sigma_{w_i}} \exp \left[ -
\frac{\Phi(z)+\frac{w^2}{2}}{\sigma_{w_i}^2} \right]  \mbox{,}
\label{e:isothermaldec}
\end{equation}
where $N$ is the number  of isothermal disks, the $i^{th}$ population having
the  vertical  velocity  dispersion  $\sigma_{w_i}$ and  the  local  density
$\rho_i(0)$.  The vertical density of a stationary population is just:
\begin{equation}
\rho(z)=\sum_{i=1}^{N} \rho_{i}(0) \exp \left[ -\Phi(z)/\sigma_{w_i}^2 \right]
=\sum_{i=1}^{N} \rho_{i}(z)  \mbox{.}
\label{e:isothermaldensite}
\end{equation}
This second  assumption is  valid as  long as the  analysed sample  covers a
restricted range of  distances close to the Galactic plane  like the bulk of
our  NGP  stars between  162\,pc  and  870\,pc.   At higher  distances,  the
potential  cannot  be  considered  as  separable  in $r$  and  $z$  and  the
inclination of the velocity ellipsoid  must be included in the analysis: see
for instance \citet{KG89} and \citet{S89}.

The vertical potential \citep{KG89} is written
\begin{equation}
\Phi(z)=2\pi G (\Sigma_0(\sqrt{z^2+D^2}-D)+\rho_{\rm eff}z^2)
\label{e:potential}
\end{equation}
a 3-parameter  ($D$, $\Sigma_0$,  $\rho_{\rm eff}$) representation  to study
the shape of the vertical  potential.  Another approach consists in building
a  stellar  disc mass  model  and  in  considering that  supplementary  mass
(unidentified mass) is proportional to the  disc mass model or to one of its
sub-component (see for instance \citet{B84,HF00}).

The total vertical  mass distribution is deduced from  $\Phi(z)$ through the
Poisson  equation and discussed  in many  papers (\citet{vH};  for extensive
references  see   \citet{CC98a,HF00},  they  also   discuss  the  horizontal
potential contribution to the dynamical estimate of the disk mass).

We  derive  the   potential  parameters  as  well  as   the  stellar  sample
distributions (through  the $\rho_i$ and $\sigma_{w_i}$) by  using a maximum
likelihood  method. The  maximum likelihood  method allows  the  fit without
binning of  data.  It allows a  non-biased estimate of  parameters even when
the number  of objects in bins would  be small and the  fluctuations are not
gaussian  and the  use of  a least  square fitting  would not  be justified.
Theoretical bases for this scheme  can be found for instance in \citet{KS73}
or in \citet{E71}.

We  set  $f^*(x_i)$ the  probability  function  of  observables $x_i$  whose
detailed expression is given below.

The logarithm of the likelihood is then defined as:
    \begin{equation} 
    \log  L = \sum_{i=1}^{n*} \log f^*(x_i) \mbox{,} 
    \label{e:fnorm}
    \end{equation}
where $n*$ is the number of red clump stars in the whole sample.
We  have separated  the fields  in subsamples  according to  distance and/or
magnitude completeness.   Then each subsample requires  a specific treatment
and  $f^*$ has  a  different  detailed algebrical  expressions.   It may  be
however written as:
$$ f^*=\frac{{\cal  M}_1+{\cal M}_2+{\cal M}_3}{{\cal  N}_1+{\cal N}_2+{\cal
N}_3}$$
where the  ${\cal M}_i$  are the distribution  function in the  volume $V_i$
expressed in the  used observable variables, and the  ${\cal N}_i$ are their
normalizations over  each volume $V_i$.   The ${\cal M}_i$ and  ${\cal N}_i$
are  explicited  in the  following  subsections  according  to the  type  of
completeness.

We have determined the model  parameters using different observables $(z, r,
b,w, m_{\rm v},\textrm{or~} M_{\rm V})$  in order to find the most efficient
and  accurate  way  to analyse  the  samples.   Model  with a  simple  limit
($z$-distance)  have a  simple algebra  and are  easier to  develop.  Models
using supplementary stars ($r$-distance  or magnitude completeness) would be
more  accurate, they  have also  more complicated  algebra.  We  compare the
numerical consistency  of the various  methods and check the  most efficient
and accurate ones.

Observations  are available  in three  volumes  ${\bf V}_k$,  the two  cones
towards the NGP ($k=1,2$) and  the local sphere ($k=3$).  The number density
of the stars is, according to the dynamical model:

\begin{eqnarray*}
{\rm d} N(x,y,z,w) = 
\begin{cases}
f(z,w)\,{\rm d}x {\rm d}y {\rm d}z {\rm d}w  \\
0 \textrm{~~if $(x,y,z,w)$ is outside {\bf V$_k$}}
\end{cases}
\end{eqnarray*}
or in  galactic coordinates:
\begin{eqnarray*}
{\rm d} N(r,l,b, w) = 
\begin{cases}
f(r \sin(b),w)\, r^2 \cos(b)\, {\rm d}r {\rm d}b {\rm d}l  {\rm d}w  \\
0 \textrm{~~if $(r,l,b,w)$ is outside  {\bf V$_k$}}
\end{cases}
\end{eqnarray*}
Due to the extended angular size  of the observed NGP fields, the dependence
on Galactic  latitude ${\it  b}$ cannot be  neglected and, by  comparison to
traditional studies, makes the algebra more tedious.
   \subsection{Local sphere}
The sample extracted from Hipparcos  data and corrected for Lutz-Kelker bias
defines a volume  that is complete in distance.  This  volume $V_{k=3}$ is a
sphere of radius $R_s=125$\,pc.

The stellar density according to variables $z$ and $w$ is modelled as:
\begin{eqnarray*}
{\rm d} {\cal N}_3(z, w)= 
\begin{cases}
\int_l \int_b {\rm d} N
= f(z,w)\, \pi\, (R_s^2-z^2)  \, {\rm d}z  {\rm d}w     \\
0 \text{~~if~}|z|>R_s 
\end{cases}
\end{eqnarray*}
We have ${\cal M}_3={\rm d} {\cal N}_3 /({\rm d}z{\rm d}w)$ and
\begin{eqnarray}
{\cal N}_3
 =  \int_{V_3} {\rm d} N  
 =  \Sigma_i \int_z  \rho_i(z) \,\pi\,(R_s^2-z^2)\, {\rm d}z \nonumber 
\end{eqnarray}

   \subsection{NGP: completeness in $z$--distances }

We have used  the largest cone samples towards the NGP  that are complete in
$z$  height above  the Galactic  plane.  For  field 1,  the volume  $V_1$ is
limited in $z$  between 226 and 391\,pc (309.4 square  degree field) and for
field 2, between 226 and 579\,pc (410.1 square degree field).

We have with $k=1,2$:
\begin{eqnarray*}
{\rm d} {\cal N}_k(z, w)=
\begin{cases}
\left( \int \left[ \frac{1}{2 (\sin b)^2}\right]_{b_{\rm low}}^{b_{\rm up}} {\rm
 d}l \right) f(z,w) \,z^2  {\rm d}z {\rm d}w \\
 0 \textrm{~~ outside of the volume {\bf V}$_k$ }
\end{cases}
\end{eqnarray*}
where  $b_{\rm low}$  and  $b_{\rm up}$  are  the lower  and upper  latitude
boundaries at fixed longitude. We have ${\cal M}_k={\rm d} {\cal N}_k /({\rm
d}z{\rm d}w)$, and
\begin{eqnarray*}
{\cal N}_k  \! = \! \int_{V_k} \! {\rm d}  N \!
 = \! \left( \int \left[ \frac{1}{2 (\sin b)^2} \right]_{b_{\rm
low}}^{b_{\rm up}}
\, {\rm d}l \right) \!
\left( \Sigma_i \int \rho_i(z)\, z^2 {\rm d}z \right)
\end{eqnarray*}
Integrals can be  separated since the $l$ and $b$ contours  do not depend on
$z$, according to the definition of the sub-samples.
   \subsection{NGP: completeness in $r$--distances }
To  increase the  number of  stars used  to constrain  the model,  we define
complete volumes in distance from  the sun, $r$.  For volume $V_1$, $r$ range
from 226 to 405\,pc and for volume $V_2$, $r$ range from 226 to 616\,pc.\\
We have with $k=1,2$:
\begin{eqnarray*}
{\rm d} {\cal N}_k(b,r, w)=
\begin{cases}
 \int_l {\rm d} N
=\Delta l(b)~f(z,w)\, r^2 \cos(b) {\rm d}r {\rm d}b {\rm d}w\\
0 \text{~~outside of the volume {\bf V}$_k$}
\end{cases}
\end{eqnarray*}
we  define ${\cal  M}_k={\rm d}{\cal  N}_k/({\rm  d}r {\rm d}b {\rm
d}w)$, and
$${\cal N}_k =\int_{Vi} {\rm d} {\cal N}_k 
=\Sigma_i \int_b  \Delta l(b) \int_r \rho_i(z)\, r^2 \cos (b) {\rm d}r{\rm d}b$$

In  the   previous  equations,  since  the  distribution   function  $f$  is
independent of $l$, the $\Delta  l(b)$ term, which is the longitude interval
at latitude  $b$ of  a circular field,  has a  closed form given  by solving
${\bf r}  \cdot {\bf r}_c =  \cos(\alpha)$, where $\alpha$ is  the radius of
the field,  ${\bf r}_c$ the unit  vector pointing towards the  center of the
field and ${\bf r}$ a unit  vector.  With $l_c$ and $b_c$ the coordinates of
the center  of the field and  $\alpha$ its radius, we  obtain the expression
valid for the field 1:
$$
\Delta          l(b)           =          2          \arccos          \left(
\frac{\cos(\alpha)-\sin(b)\sin(b_c)}{\cos(b)\cos(b_c)} \right) \, \mbox{.}
\label{e:deltal1}
$$
For the field  2, $\Delta l(b)$ is $2\pi$ since the  field is centred toward
the Galactic pole,  unless the latitude $b$ corresponds  to the removed area
of the Coma  Berenices cluster or to the field  1 overlapping region.  Since
these removed region are circular, we have:
$$
\Delta l(b) = 2 \pi - \Delta_{coma}l(b) -\Delta_{field\,1}l(b)\,\,\mbox{.}
\label{e:deltal2}
$$
   \subsection{NGP: apparent magnitude completeness}

Since the NGP  samples have been selected by apparent  magnitude, it is more
efficient, in  terms of  the number  of stars used,  to keep  this criterion.
Modelling the  distribution of apparent  magnitudes of selected  clump stars
allows the use of 201 stars from 160 to 870\,pc from the Galactic plane.\\
With $k=1,2$, we have:
\begin{eqnarray*}
{\cal M}_k = {\rm d}{\cal N}_k (b, m_{\rm v}, w)/ ({\rm d}b {\rm
d}m_{\rm v} {\rm d}w )=  \\
~~
\begin{cases}
=\Delta l(b) \int_{r=0}^{\infty} f(r \sin(b),w) \Phi(M_{\rm V})\,r^2 \cos (b)\,{\rm d}r  \\
\textrm{or~} = 0 \textrm{~~if~} m_{\rm v}\, \textrm{is outside the limits given in Table\,\ref{t:samples}}
\end{cases}
\end{eqnarray*}
where $\Delta l(b)$ has been previously defined.  We have also:
$${\cal N}_k \!
=\!\Sigma_i \! \int_b \! \Delta l(b) \int \!\!  \int 
\rho_i(r \sin(b))\Phi(M_{\rm V})\,r^2
\cos (b)\, {\rm d}r{\rm d}b{\rm d}m_{\rm v}$$
  \subsection{NGP:  apparent  magnitude  completeness  (with  absolute
  magnitudes and distances)}
We use the  same selection criterion in apparent  magnitudes, with 200 stars
in the range 200--750\,pc from the Galactic plane.\\
Here we  model both the  distributions of absolute magnitudes  and distances
expecting a more constraining use of the observables $M_{\rm V}$\\
We have with $k=1,2$
\begin{eqnarray*}
{\cal M}_k= {\rm d}{\cal N}_k (z, b, w, M_{\rm V})/({\rm d}r{\rm
d}b{\rm d}w{\rm d}M_{\rm V})=   \\
~~
\begin{cases}
=\Delta l(b) f(z,w) \Phi(M_{\rm V}) r^2 \cos(b)    \\
\textrm{or~} = 0 \textrm{~~if~} m_{\rm v} \,
\textrm{is outside the limits given in Table\,\ref{t:samples}}
\end{cases}
\end{eqnarray*}
and we have:
\begin{eqnarray*}
{\cal N}_k\! =\!   
\Sigma_i\!\!\int_b\!\! \Delta l(b)\!\!\int_r\,\int_{M_{\rm V}}\!\!\!\rho_i(r \sin
(b)) \Phi(M_{\rm V})\, r^2\!\! \cos(b)\, {\rm d}r {\rm d}M_{\rm V} {\rm d}b
\end{eqnarray*}
%
%
\section{Results}
\paragraph{Modelling.}

The modelling of the samples consists in reproducing the distribution of the
observed  apparent magnitudes, absolute  magnitudes, distances  and vertical
velocities by  adjusting the three vertical potential  parameters (the total
surface mass  density $\Sigma_0$,  the mass scale  height $D$, and  the halo
local effective  mass density $\rho_{\textrm{eff}}$) as well  as the stellar
density and velocity distributions (through the velocity dispersions and the
respective  contribution  of each  isothermal  component).   Using the  four
methods previously described, the model parameters are adjusted by a maximum
likelihood.  The  maximum  likelihood  method  avoids  bias  that  would  be
introduced by a  binning of data and  by the small size of  our samples.  As
expected, the accuracy  is better with methods using  more stars and/or more
distant  stars,  while the  comparison  of  methods  allows to  check  their
respective robustness.  Following discussions are based on the results given
by methods number 3 and 4.

\paragraph{$\rho_{\textrm{eff}}$.}

The  parameter  $\rho_{\textrm{eff}}$  is  introduced  to  model  the  local
contribution  of  a spherical  massive  halo.  A  realistic range  for  this
parameter is 0  to 0.02 M$_{\odot}${\hspace{0.25em}}pc$^{- 3}$ \citep{KG89}.
However, this  halo contribution to  the vertical potential is  quadratic in
$z$  (at  small $z$)  and  cannot be  clearly  distinguished  from the  disk
contribution (the  remaining terms in  Eq.  \ref{e:potential}) that  is also
quadratic for small $z$.  For these reasons, $\rho_{\textrm{eff}}$ is set at
$0.01${\hspace{0.25em}}M$_{\odot}${\hspace{0.25em}}pc$^{- 3}$.

\paragraph{Adjusted parameters.}

The best  decomposition of the  observed stellar distributions  in kinematic
components,  using  methods  3  or  4, is  obtained  with  three  isothermal
components  with densities  (22\%, 59\%  and 19\%)  and  respective vertical
velocity  dispersions  $\sigma_w$  (8.5,  11.4 and  31.4)  km\,s$^{-1}$.   A
two-component decomposition gives nearly the same likelihood with respective
densities (78\% and 22\%) and  vertical velocity dispersions (12.0 and 30.4)
km\,s$^{-1}$.  This  result can be  compared to the  kinematic decomposition
obtained in paper  I \citep{SBS02} using the 3D  velocity components and the
metallicity.

\begin{figure}[hbtp]
\centering
\includegraphics[width=6.5cm]{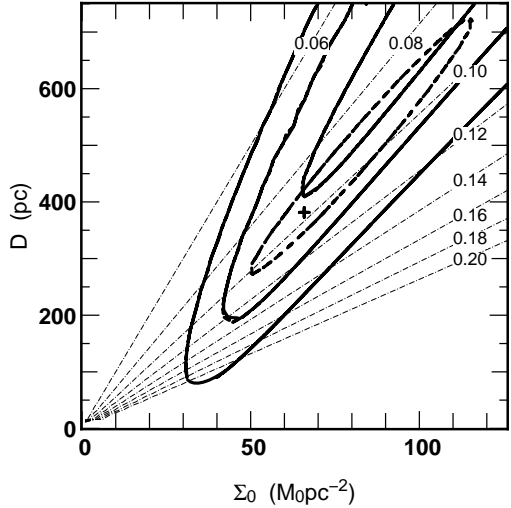}
\caption{1,  2  and  3--$\sigma$  error  contours  for  $\Sigma_0$  and  $D$
solutions obtained with method 4  (continuous curves). Using also the recent
Oort  limit determination  0.0102\,M$_{\odot}$\,pc$^{-3}$  from \citep{HF00}
gives   better  constraints   (1--$\sigma$  error   level:   dashed  curve).
Dashed--dotted  lines   show  solutions  for  adopted   local  mass  density
$\rho_{\rm   total}$  values   between   0.06  to   0.2\,$\mathrm{M}_{\odot}
\mathrm{pc}^{-3}$.}
\label{fig:isolevel}
\end{figure}

The maximum  of likelihood  gives also the  following result: with  method 3
($D$, $\Sigma_0$)=  (492\,pc, 72  M$_{\odot}$\,pc$^{-2}$) and with  method 4
($D$, $\Sigma_0$)= (1500\,pc,  193 M$_{\odot}$\,pc$^{-2}$).  At a 1-$\sigma$
error level  methods 3 and  4 give only  a lower limit for  $D$ respectively
227\,pc and 400\,pc and no  upper limit.  The $\Sigma_0$ values are strongly
correlated with $D$ (see  Figure \ref{fig:isolevel}).  We must consider that
our sample  puts lower limits to  ($D$, $\Sigma_0$) parameters  and a sample
with higher $z$-distances would be needed to put upper limits.

If $D$ were  known independently, the relative accuracy  on $\Sigma_0$ would
be  12   \%,  while  considering  the  seven   model  parameters  (including
dispersions  and relative  density of  the two  stellar components,  $D$ and
$\Sigma_0$), we find a large  correlation between the two parameters $D$ and
$\Sigma_0$.   We may  note that  we  obtain the  first direct  observational
constraint  on the  thickness of  the  total disk  mass.  We  also obtain  a
constraint       on      the       total       local      mass       density
$\rho_{\rm{total}}$(z=0)=0.08$\pm$0.01 M$_{\odot}$\,pc$^{-3}$ (see below).

\paragraph{Comment about $D$.}

In  some  previous   vertical  Galactic  force  determinations,  simplifying
assumptions have  been adopted  concerning the exact  shape of  the vertical
mass distribution.  For instance,  \citet{KG89} modelled the potential using
Eq.~\ref{e:potential}, but  assuming $D$\,=\,180\,pc. They  argue that their
result is not sensitive to a small change on $D$ since the majority of their
sample  ranges beyond  500\,pc from  the  Galactic plane.   However since  a
significant  fraction of  their sample  ranges between  200 and  500\,pc, we
suspect,  that with a  drastic change  of $D$  at $\sim$500\,pc,  their data
would   be  fitted  with   a  higher   $\Sigma_0$  value.    Other  previous
determinations of  the total  surface mass density,  based on  one parameter
models for the potential, could suffer a similar bias.

\begin{table}[hbtp]
\caption{\label{t:results}  $\Sigma_0$ and $D$ solutions for
the potential within 1--$\sigma$ error and associated total local mass density
and column density.}
\begin{center}
\begin{tabular}{c c c c c c c c c c c}
\hline
\hline
$\Sigma_0$ & $\mathrm{M}_{\odot}\mathrm{pc}^{-2}$ & 64. & 90. & 193. \\
$D$ & pc &  400 & 637 & 1500 \\
$\rho_{\rm eff}$ & $\mathrm{M}_{\odot}\mathrm{pc}^{-3}$ &  0.01   & 0.01 & 0.01 \\
$\rho_{\rm total}(0)$ & $\mathrm{M}_{\odot}\mathrm{pc}^{-3}$ & 0.090 & 0.081  & 0.074 \\
$\Sigma (800$\,pc) & $\mathrm{M}_{\odot}\mathrm{pc}^{-2}$ & 73.2 & 86.4 &106.8 \\
$\Sigma (1.1$\,kpc) & $\mathrm{M}_{\odot}\mathrm{pc}^{-2}$ & 82.2 &99.9 &136.1 \\
\hline
\end{tabular}
\end{center}
\end{table} 

\begin{figure*}
\centering
\includegraphics[width=6.5cm]{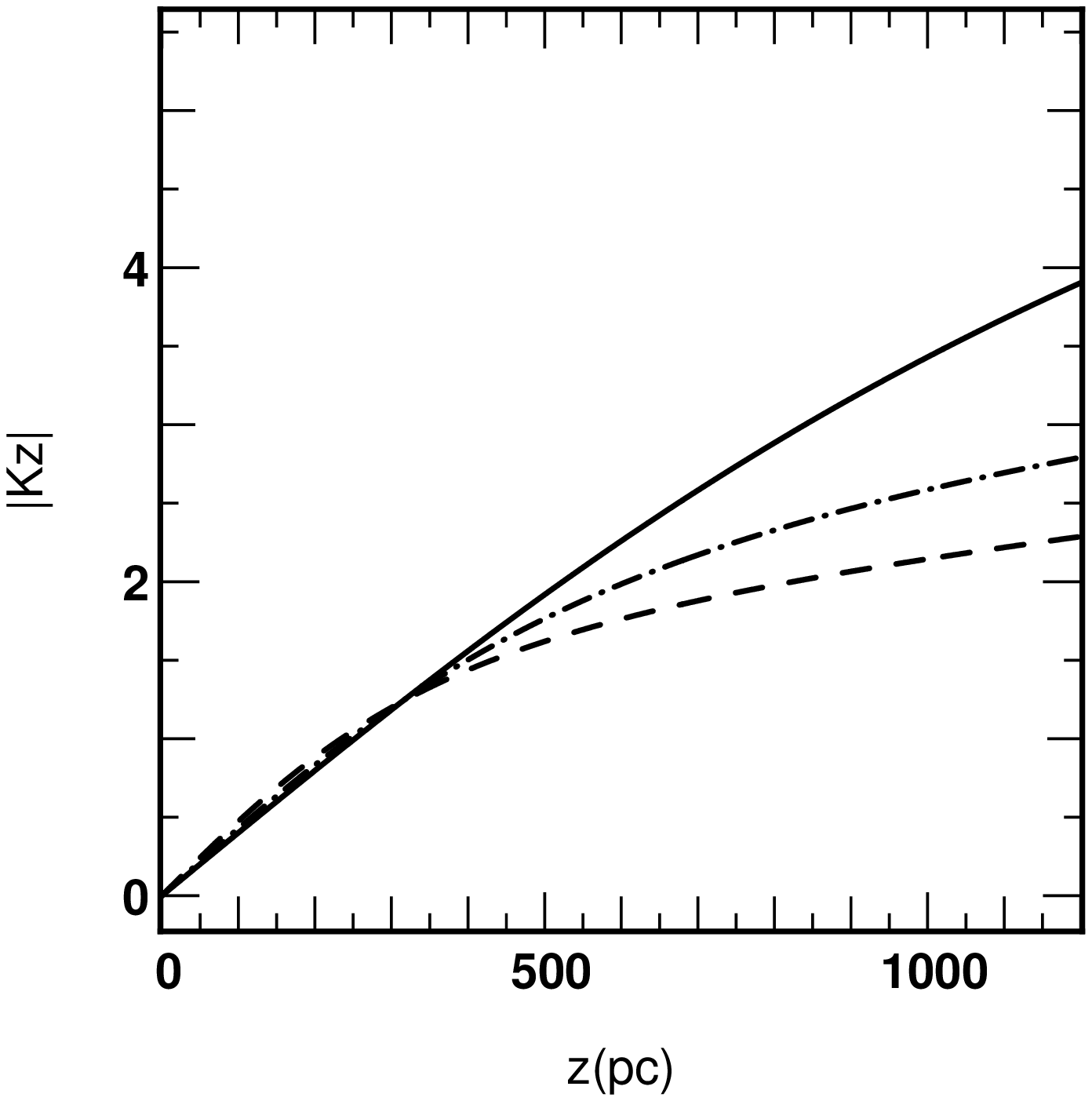}
\includegraphics[width=6.5cm]{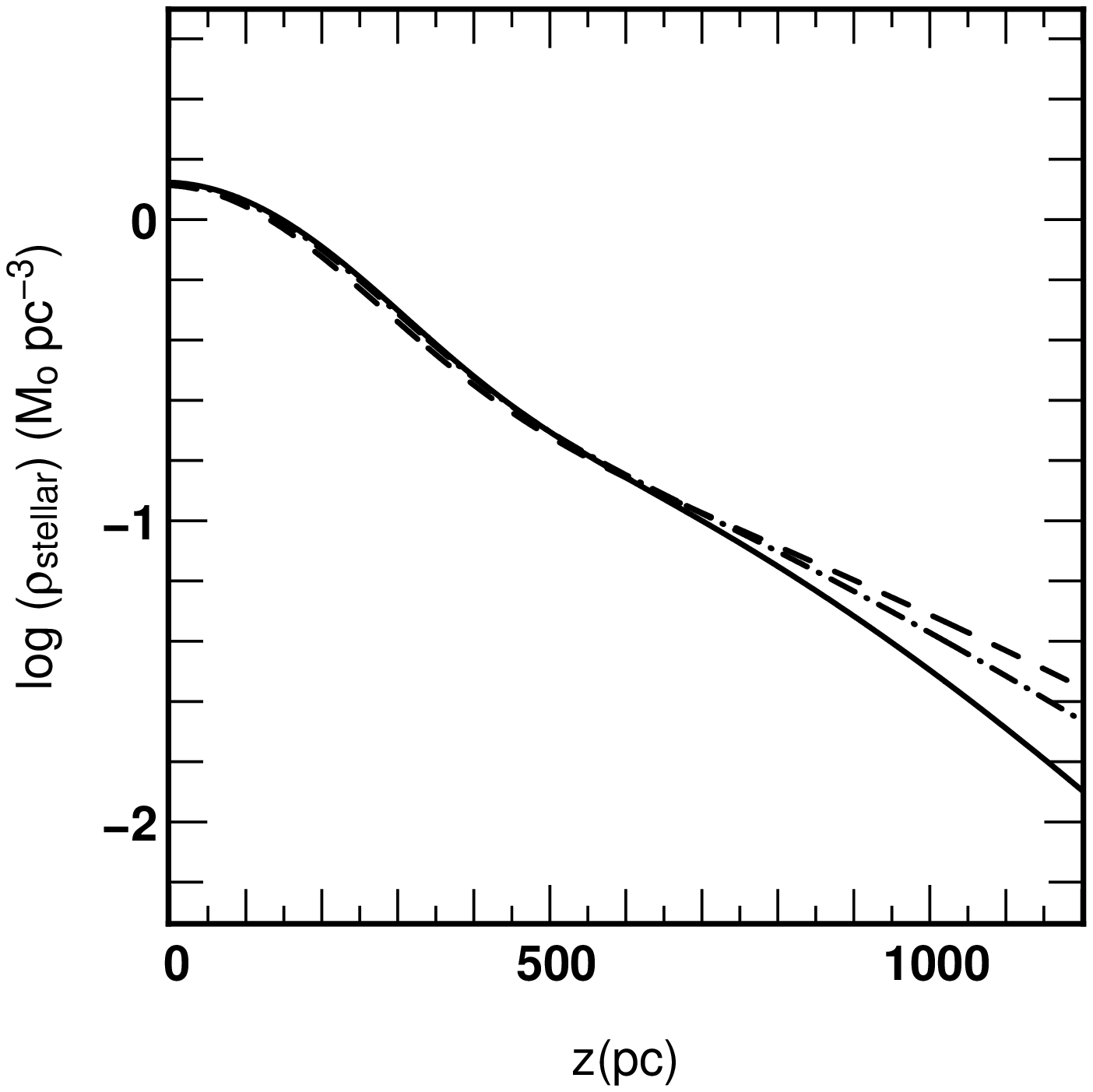}
\caption{Left  : vertical  force in  (km/s)$^2$/pc for  the  three solutions
within 1--$\sigma$  presented in Tab.~2. Right :  associated stellar density
in log (M$_\odot$ pc$^{-3}$).}
\label{fig:1sigma}
\end{figure*}

\paragraph{Total local mass density $\rho_{\rm{total}}$(z=0).}

A set of $\Sigma_0$ and $D$  solutions within 1--$\sigma$ error are given in
Table~2   with  their   corresponding  total   local  mass   volume  density
$\rho_{\rm{total}} ( z = 0 )$ (according to Eq.  \ref{e:potential} the local
mass density  is given  by $\rho_{\rm{total}}(z=0) =  \Sigma_0 /  ( 2 D  ) +
\rho_{\rm{eff}}$).  We also give the  total mass density within 800\,pc from
the  Galactic  plane  including  the  halo  contribution  parametrized  with
$\rho_{\rm eff}$ (our surface mass  density determination applies only up to
$\sim$\,800\,pc, the limit  of our sample).  The acceptable  range of values
for $\Sigma_0$ and $D$ is large,  resulting in a wide range of possible mass
distributions.     This   can    be   seen    on   the    left    panel   of
figure~\ref{fig:1sigma} where are plotted the $K_z$ solutions given in Tab~2
(recall that $K_z$ is proportional to $\Sigma(z)$) .

To disentangle  the different  solutions, we need  more observations  in the
range  750 pc  to  1.1  kpc.  This  is  illustrated on  the  right panel  of
Fig.~\ref{fig:1sigma} by  the predicted clump giant  density (the observable
quantity),   where    we   can   see   that   the    solutions   quoted   in
Tab.~\ref{t:results} differ significantly above 750 pc.

We may remark  that the total local mass density $\rho_{\rm{total}}  ( z = 0
)$ given in Table~2  ranges below 0.10 $\mathrm{M}_{\odot} \mathrm{pc}^{-3}$
and  is in  agreement with  recent and  independent determinations  based on
Hipparcos data, compatible with local mass density of known matter.

\paragraph{Additional constraint from $\rho_{\rm total}(z=0)$:  the Oort limit.}

In Table~2, the  different solutions differ by their  predicted total volume
mass density, from 0.07 to 0.09\, M$_{\odot}${\hspace{0.25em}}pc$^{- 2}$ but
remain in  agreement with the recent  measures of $\rho_{\rm{total}}$($z$=0)
based on Hipparcos data  {\citep{CC98a, CC98b,Ph98,HF00}} ranging from 0.076
to 0.102 $\mathrm{M}_{\odot} \mathrm{pc}^{-3}$.

 The data set of clump giants that  we have used does not allow to constrain
more efficiently  the local volume  mass density $\rho_{\rm  total} ( z  = 0
)$. Most of the constraints would come from the local sample within 125\,pc,
but  due to  the high  velocity dispersions  of the  red giant  stars, their
density distribution is nearly uniform within 125\,pc and the bending of the
vertical  density  distribution due  to  the  potential  cannot be  measured
significantly.  We notice  that the analysis of red clump  stars in a sphere
of 250\,pc would help to constrain  the local volume density since they have
distances  measured by  Hipparcos but  they  have no  measured or  published
radial velocities.  This would give a new independent determination, however
the  amplitude  of the  Lutz-Kelker  or Malmquist  bias  will  be large  and
certainly difficult to model properly.

\begin{figure*}
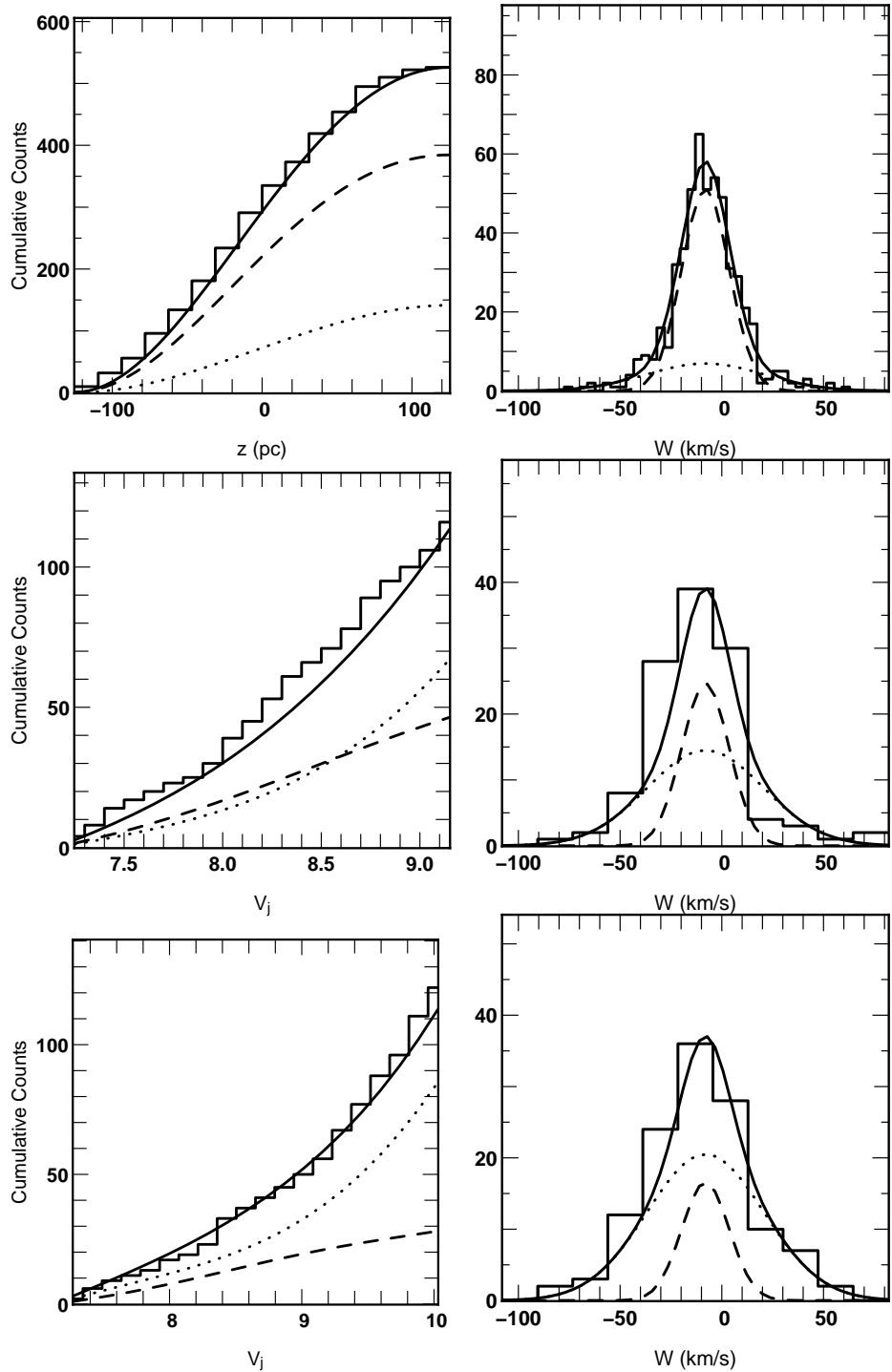

\centering
\includegraphics[width=6cm]{zcum_hip.ps}
\includegraphics[width=6cm]{W_hip.ps}
\includegraphics[width=6cm]{Vjcum_r15.ps}
\includegraphics[width=6cm]{W_r15.ps}
\includegraphics[width=6cm]{Vjcum_r10.ps}
\includegraphics[width=6cm]{W_r10.ps}
\caption{Predicted counts for  our best fit model of  the vertical potential
(solid  line)  compared to  the  observed  distribution  (histogram). Top  :
Hipparcos sample.   Middle : NGP sample  of radius 15$^\circ$.  Bottom : NGP
sample of radius 10$^\circ$.  Left : cumulative distribution of $z$ distance
(Hipparcos sample) or $V_j$ (NGP  samples). Right : distribution of vertical
velocities.   The dashed  and  dotted  lines are  the  contribution of  each
isothermal population.}
\label{fig:fitdata}
\end{figure*}

So  we consider  the recent  results based  on Hipparcos  data of  the local
volume    mass   density    $\rho_{\rm{total}}$($z$=0).     Differences   on
$\rho_{\rm{total}}$($z$=0) between \citet{CC98a} and \citet{HF00} are within
1--$\sigma$  and   are  explained  by  \citet{HF00}  as   due  to  different
assumptions on  the exact  shape of the  potential.  \citet{CC98a}  assume a
quadratic potential  close to the Galactic plane  while \citet{HF00} suppose
that the vertical potential is proportional to a more realistic model of the
vertical   stellar  disk  distribution.    Admitting  this   more  realistic
hypothesis for the  shape of the potential, we  consider that Hipparcos data
gives  $\rho_{\rm{total}}  (z=0)$  \,  =  \,  0.102 \,  $\pm$  \,  0.010  \,
M$_{\odot}$  {\hspace{0.25em}}pc$^{-3}$   \citep{HF00}  and  combining  this
result  and our result  obtained from  distant red  clump giants,  we deduce
stronger  constraints  on  $\Sigma_0$  and  $D$  (Figure~\ref{fig:isolevel}:
dashed contour draws the 1-$\sigma$ limit).  We deduce that the scale height
$D$  is relatively  large,  $390^{+330}_{-120}$\,pc, as  well  as the  total
surface  mass  density  $\Sigma_0$\,=\,67\,$^{+47}_{-18}\,\mathrm{M}_{\odot}
\mathrm{pc}^{-2}$,      and     the      mass     within      800\,pc     is
$\Sigma$($|z|<$800\,$\rm{pc}       )$\,=\,76$_{-12}^{+25}\,\mathrm{M}_{\odot}
\mathrm{pc}^{-2}$.       The      quantity     $\Sigma$($|z|<$1100\,$\rm{pc}
)$\,=\,85$_{-13}^{+32}\,\mathrm{M}_{\odot} \mathrm{pc}^{-2}$ can be directly
compared      to     the      result      obtained     by      \cite{KG91}~:
$\Sigma$($|z|<$1100\,$\rm{pc}              )$\,=\,71$\pm6\,\mathrm{M}_{\odot}
\mathrm{pc}^{-2}$. The two results are compatible in the 1$\sigma$ limit but
our  error bars  are  very large  due to  the  poor constraint  on $D$.  The
agreement between the  maximum likelihood solution and the  data is shown in
figure~\ref{fig:fitdata}. 
The  NGP sample  exhibits an  offset of the  $w$ velocity  towards more
negative values than  the expected -7 km/s due to  solar motion. This offset
is visible at  all distances, however its increase with  $z$ is doubtful. We
have  to consider  that a  negative offset  is a  real feature  in  the data
because our  $w$ velocities have an accuracy  better than 1 km/s  due to the
contribution  of the  ELODIE  radial velocities  by  more than  94\% in  the
considered  direction. Nevertheless,  this  offset is  not  constant and  is
therefore not  accounted for in  the modeling. Table~\ref{t:wvel}  shows the
mean  value of  $w$, standard  deviation and  standard error  on mean  W for
various intervals in $z$.

\begin{table*}
\center
\caption{\label{t:wvel}
$w$ behaviour as a function of $z$.}
\begin{tabular}{c c c c c c}\\
\hline
\hline
\multicolumn{2}{c}{$z$ interval} & number of stars & $<w>$ & standard deviation &
mean error \\
\hline
-125 &   0 & 252 & -7.8 & 18.4 & 1.2 \\
   0 & 125 & 270 & -7.9 & 17.1 & 1.1 \\
 125 & 250 &  30 & -13.8 & 11.7 & 2.1 \\
 250 & 375 &  46 & -12.4 & 22.5 & 3.3 \\
 375 & 500 &  56 & -7.7 & 27.9 & 3.7 \\
 500 & 625 &  64 & -15.5 & 26.4 & 3.3 \\
 625 & 750 &  52 & -9.3 & 31.6 & 4.4 \\
\hline
\end{tabular}
\end{table*}

\section{Discussion}

With samples of giants in the solar neighbourhood up to distances of 800\,pc
toward the NGP,  we solve for the vertical Galactic  potential and the total
disk surface mass density at the solar Galactic position.

We   find  a   vertical   potential  $\Phi(z)$   compatible   at  high   $z$
($\sim$\,400--800\,pc)  with previous  estimates \citep{KG91,FF94}.   We are
also able  to measure for  the first time,  the thickness of the  total disk
mass  distribution, and  find  a characteristic  height  $D$=390\,pc with  a
1--$\sigma$ range from 271\,pc to 720\,pc.

\begin{figure}[hbtp]
\centering
\includegraphics[width=8.5cm]{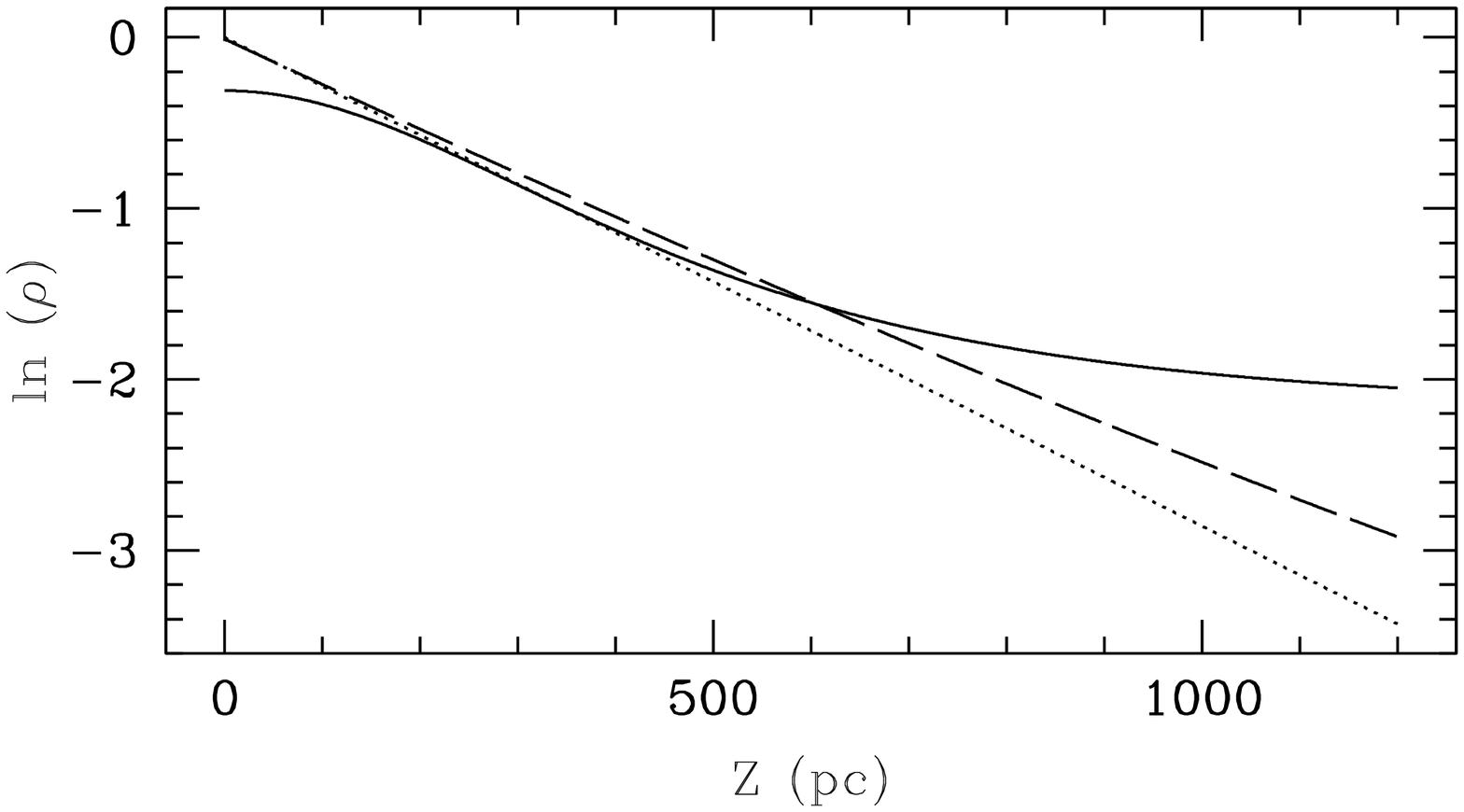}
\caption{Vertical distribution  of the volume  mass density in  the Galactic
disk  deduced from our  dynamical determination  (continuous line).   A disk
vertically  exponential  with a  350\,pc  scale  height  (dotted line),  and
complemented with a  thick disk (scale height 750\,pc,  and relative density
of 15 \% at $z$=0 (dashed line).}
\label{fig:diskdensity}
\end{figure}

This scale height $D$ corresponds to  a 350\,pc scale height of a vertically
exponential density law\footnote {The validity of exponential density models
to represent the vertical structure  of the disk is questioned and discussed
by  \citet{H97a,H97b} and  he explains  some of  the  systematic discrepancy
between  authors  and  that  at  low distances  (below  500\,pc),  a  single
exponential  is  inadequate  under  any reasonable  scenario.   A  practical
consequence is that  an exponential fitting to star counts  is valid only in
restricted range of distances and must  not be extrapolated to $z$=0.  }  in
the range of $z$-height  200 to 500\,pc.  Figure~\ref{fig:diskdensity} shows
also the  resulting total mass  density from Eq.  \ref{e:potential}  and the
stellar density for two exponential  disks (old and thick disks respectively
with 350\,pc  and 750\,pc  scale heights  and a relative  density of  15 per
cent).   This measure  of the  mass scale  height is  in agreement  with the
recent  determination of the  stellar disk  scale height,  h$_z$=330\,pc, by
\citet{C01} based  on SDSS star  counts and compatible with  lower estimates
from  star   counts  (for  instance   $260$\,$\pm$\,50\,pc  by  \citet{O96},
250\,$\pm$\,60\,pc  by \citet{V00}  or  260\,$\pm\,90\,$pc by  \citet{S02}).
Below $z$=600\,pc, the contribution of the stellar thick disk remains always
much smaller than the old  disk, and beyond 600\,pc its contribution remains
smaller than the dark halo (see Figure~\ref{fig:diskdensity}).

We find for the surface mass  density within 800\,pc from the Galactic plane
a  value  of  $\Sigma$(800\,pc)=\,76\,$\mathrm{M}_{\odot}  \mathrm{pc}^{-2}$
(this value includes the contribution from known or ``seen'' matter -stellar
and gas- and a contribution from a  round massive dark halo) and we find for
the  total disk  surface  mass density  $\Sigma_0$=\,67\,$\mathrm{M}_{\odot}
\mathrm{pc}^{-2}$.  This disk  with a round dark halo  (with a local density
$\sim$0.01$\mathrm{M}_{\odot}  \mathrm{pc}^{-3}$) may  produce  the observed
flat rotation  curve.  Modellings  with similar parameters  can be  found in
\citet{KG91} and \citet{CC98a}.

Our     1-$\sigma$     error     upper     limit    allows     a     maximum
$\Sigma_0$\,=\,114\,$\mathrm{M}_{\odot} \mathrm{pc}^{-2}$ and a scale height
of  720\,pc.   It  is  about the  upper  limit  $\sim100\,\mathrm{M}_{\odot}
\mathrm{pc}^{-2}$ putting all  the Galactic mass in the  disk (maximal disk)
and still  explaining the observed Galactic rotation  curve \citep{S97}.  It
would  also imply that  the dark  matter is  within a  disk of  about 1\,kpc
thickness.  Such a flat dark  matter has been rejected by some observational
evidences: for instance on the basis of the observed coupling in the stellar
3D velocity  distribution in the solar neighbourhood  \citep{B99} implying a
more or less  round dark halo, or also on  the non-precessing circular orbit
of the Sagittarius tail around our Galaxy \citep{I01} implying also that the
halo is most likely spherical.

We conclude that the local volume  mass density, the surface mass density of
the disk are in agreement with our current knowledge of the known volume and
surface density from gas and stellar components.  Moreover, the thickness of
the disk mass  density distribution is compatible with  the thickness of the
stellar old disk.


\begin{acknowledgements}
This  research  has made  use  of  the  SIMBAD and  VIZIER  databases,
operated at CDS, Strasbourg,  France.

This paper  is based  on data from  the ESA {\it  Hipparcos} satellite
(Hipparcos and Tycho-II catalogues).
\end{acknowledgements}
%
%
%
\bibliographystyle{aa}

\begin{thebibliography}{}
\expandafter\ifx\csname natexlab\endcsname\relax\def\natexlab#1{#1}\fi

\bibitem[{{Arenou} \& {Luri}(1999)}]{Ar99}
{Arenou}, F. \& {Luri}, X. 1999, in ASP Conf. Ser. 167: Harmonizing Cosmic
  Distance Scales in a Post-Hipparcos Era, 13--32

\bibitem[{{Bahcall}(1984)}]{B84}
{Bahcall}, J. N. 1984, \apj, 276, 156

\bibitem[{{Barbier--Brossat} \& {Figon}(2000)}]{BB00}
{Barbier--Brossat}, M. \& {Figon}, P. 2000, \aaps, 142, 217

\bibitem[{{Bienaym\'e} (1999)}]{B99}
Bienaym\'e, O. 1999, \aap, 341, 86

\bibitem[{{Chen et al.} (2001)}]{C01}
Chen, B., Stoughton, C., Smith, J.\,A. et al. 2001, \apj, 553, 184



\bibitem[{{Chabrier} (2001)}]{Cha01}
Chabrier, G. 2001, \apj, 554, 1274

\bibitem[{{Chabrier} (2002)}]{Cha02}
Chabrier, G. 2002, \apj, 567, 304

\bibitem[{{Cr\'ez\'e} {et~al.}(1998a){Cr\'ez\'e}, {Chereul}, {Bienaym\'e}, \&
  {Pichon}}]{CC98a}
{Cr\'ez\'e}, M., {Chereul}, E., {Bienaym\'e}, O., {Pichon}, C. 1998, \aap, 
  329, 920

\bibitem[{{Cr\'ez\'e} {et~al.}(1998b)}]{CC98b}
{Cr\'ez\'e}, M., {Chereul}, E., {Bienaym\'e}, O., {Pichon}, C. 1998, 
 Proc. of the ESA Symp. ``Hipparcos - Venice 97'', ESA SP-402, 669


\bibitem[{{Eadie et al.}(1971)}]{E71}
Eadie, W.T., Drijard, D., James, F.E., Roos, M., \& Sadoulet, B. 1971,
{Statistical Methods in Experimental Physics}, Ch. 8, North Holland (Amsterdam)

\bibitem[{{ESA} (1997)}]{HIP}
ESA 1997, {The Hipparcos and Tycho Catalogues}, (Noordwijk) Series: ESA--SP 1200

\bibitem[{{Flynn \& Fuchs}(1994)}]{FF94}
Flynn, C., Fuchs, B. 1994, \mnras, 270, 471

\bibitem[{{Girardi} {et~al.}(1998){Girardi}, {Groenewegen}, {Weiss}, \&
  {Salaris}}]{Gi98}
{Girardi}, L., {Groenewegen}, M.\,A.\,T., {Weiss}, A.,  {Salaris}, M. 1998,
  \mnras, 301, 149


\bibitem[{{Haywood et al.} (1997a)}]{H97a}
Haywood, M., Robin, A.C., Cr\'ez\'e, M. 1997, \aap, 320, 428

\bibitem[{{Haywood et al.} (1997b)}]{H97b}
Haywood, M., Robin, A.C., Cr\'ez\'e, M. 1997, \aap, 320, 440

\bibitem[{{Holmberg} \& {Flynn}(2000)}]{HF00}
{Holmberg}, J.,  {Flynn}, C. 2000 \mnras,  313, 209

\bibitem[{{Ibata et al.} (2001)}]{I01}
Ibata, R., Geraint, F., Irwin, M., Totten, E., Quinn, T. 2001 \apj, 551, 294

\bibitem[{{Kapteyn}(1922)}]{K22}
Kapteyn, J.C. 1922, \apj, 55, 302 

\bibitem[{{Katz} {et~al.}(1998){Katz}, {Soubiran}, {Cayrel}, {Adda}, \&
  {Cautain}}]{KS98}
{Katz}, D., {Soubiran}, C., {Cayrel}, R., {Adda}, M., \& {Cautain}, R. 1998,
  \aap, 338, 151

\bibitem[{{Kendall \& Stuart}(1973)}]{KS73}
Kendall, M.G., Stuart, A. 1973, { The Advanced Theory of Statistics,
Vol. 2}, Ch. 18, Ed. Griffin (London) 

\bibitem[{{Kerr \& Lynden-Bell}(1986)}]{KL86}
Kerr, F.J., Lynden-Bell, D. 1986, \mnras, 221, 1023

\bibitem[{{Kuijken}(1995)}]{K95}
Kuijken, K. 1995, Stellar populations, IAU Symp. 164, 195 

\bibitem[Kuijken and Gilmore (1991)]{KG91}
Kuijken K., Gilmore G., 1991, \mnras, ~313, 209

\bibitem[Kuijken and Gilmore (1989)]{KG89}
Kuijken K., Gilmore G., 1989, \mnras, ~239, 605

\bibitem[{{Lutz} \& {Kelker}(1973)}]{LK73}
{Lutz}, T.~E., {Kelker}, D.~H. 1973, \pasp, 85, 573

\bibitem[{{Ojha et al.}(1996)}]{O96}
Ojha, D.\,K, Bienaym\'e, O., Robin, A., Cr\'ez\'e, M., Mohan, V 1996, \aap, 311, 456

\bibitem[{{Oort}(1932)}]{O32}
Oort, J.H. 1932, BAN, 6, 249 

\bibitem[{{Pham}(1998)}]{Ph98}
{Pham}, H.-A. 1998,  Proc. of the ESA Symp. ``Hipparcos - Venice 97'', ESA SP-402,  559

\bibitem[{{Prugniel} \& {Soubiran}(2001)}]{PS01}
{Prugniel}, P. \& {Soubiran}, C. 2001, \aap, 369, 1048

\bibitem[{{Soubiran et al}(2002)}]{SBS02}
Soubiran, C., Bienaym\'e, O., Siebert, A. 2002, \aap, submitted (Paper I)

\bibitem[{{Sackett}(1997)}]{S97}
Sackett, P. 1997 \apj, 483, 103 

\bibitem[{{Sohn}(2002)}]{S02}
Sohn, Y-J 2002,  Journ. Astr. Sp. Sci. , 19, 19

\bibitem[{{Statler}(1989)}]{S89}
Statler, T.S.   1989, \apj, 344, 217

\bibitem[{{Turon--Lacarrieu} \& {Cr\'ez\'e}(1977)}]{TLC77}
{Turon--Lacarrieu}, C. \& {Cr\'ez\'e}, M. 1977, \aap, 56, 273

\bibitem[{{Vallenari et al.}(2000)}]{V00}
Vallenari, A., Bertelli, G., Schmidtobreick, L. 2000, \aap, 361, 73

\bibitem[{von Hoerner}(1960)]{vH}
von Hoerner, S. 1960, Fortschr. Phys., 8, 191

\end{thebibliography}

\end{document}